\documentclass[english]{llncs}
\usepackage{lmodern}
\usepackage[T1]{fontenc}
\usepackage[latin9]{inputenc}
\usepackage{array}
\usepackage{multirow}
\usepackage{amstext}
\usepackage{amssymb}
\usepackage{graphicx}

\makeatletter

\providecommand{\tabularnewline}{\\}

\usepackage{graphicx}
\usepackage[font={small}]{caption}

\makeatother

\usepackage{babel}
\begin{document}

\title{Jointly Discriminative and Generative Recurrent Neural Networks for
Learning from fMRI}

\author{Nicha C. Dvornek\textsuperscript{1,2}, Xiaoxiao Li\textsuperscript{2},
Juntang Zhuang\textsuperscript{2}, James S. Duncan\textsuperscript{1,2,3,4}}

\institute{\textsuperscript{1}Department of Radiology \& Biomedical Imaging,
Yale School of Medicine, New Haven, CT, USA\\
\textsuperscript{2}Department of Biomedical Engineering, \textsuperscript{3}Department
of Electrical Engineering, and \textsuperscript{4}Department of Statistics
and Data Science, Yale University, New Haven, CT, USA}
\maketitle
\begin{abstract}
Recurrent neural networks (RNNs) were designed for dealing with time-series
data and have recently been used for creating predictive models from
functional magnetic resonance imaging (fMRI) data. However, gathering
large fMRI datasets for learning is a difficult task. Furthermore,
network interpretability is unclear. To address these issues, we utilize
multitask learning and design a novel RNN-based model that learns
to discriminate between classes while simultaneously learning to generate
the fMRI time-series data. Employing the long short-term memory (LSTM)
structure, we develop a discriminative model based on the hidden state
and a generative model based on the cell state. The addition of the
generative model constrains the network to learn functional communities
represented by the LSTM nodes that are both consistent with the data
generation as well as useful for the classification task. We apply
our approach to the classification of subjects with autism vs. healthy
controls using several datasets from the Autism Brain Imaging Data
Exchange. Experiments show that our jointly discriminative and generative
model improves classification learning while also producing robust
and meaningful functional communities for better model understanding.
\end{abstract}

\section{Introduction}

Functional magnetic resonance imaging (fMRI) has become an important
tool for investigating neurological disorders and diseases. In addition,
machine learning has begun to play a large role, in which classification
models are learned and interpreted to discover potential fMRI biomarkers
for disease. Traditional approaches for building classification models
from resting-state fMRI first parcellate the brain into a number of
regions of interest (ROIs) and use functional connectivity between
the ROIs as inputs to a classification algorithm \cite{Abraham2017}.
Recently with the advent of deep learning, temporal inputs based on
the time-series data combined with recurrent neural network (RNN)
models have been explored for predicting from fMRI \cite{Dvornek2017,Gueclue2017,Li2018a}.
Such RNN models are attractive for processing fMRI as they were designed
for dealing with sequential data. However, the large sample sizes
required for effective deep learning are difficult to gather for fMRI
data, particularly for many different patient populations or types
of studies. 

One way to handle the limited data problem is to apply multitask learning
\cite{Caruana1997}. The idea in multitask learning is that shared
information across related tasks is jointly learned in order to improve
the learning of each individual task. For a classification task based
on fMRI data, e.g., distinguish subjects with a given disease from
healthy individuals, the amount of labeled data is often limited.
Thus, we propose to apply multitask learning to improve the learning
of a target discriminative task by jointly learning an auxiliary generative
model for the fMRI data, which does not require any annotation. Moreover,
simultaneous learning of the generative model will assist in interpreting
the discriminative model. 

Specifically, we propose to jointly learn a discriminative task while
also learning to generate the input fMRI time-series by using an RNN
with long short-term memory (LSTM). Generative RNN models have been
extensively used in natural language processing, e.g., for text generation
\cite{Graves2014}, but application to the medical imaging field has
been limited. Furthermore, multitask learning with discriminative
and generative components have been combined in many different neural
network architectures, notably generative adversarial networks, but
such a joint learning approach utilizing the RNN framework has only
begun to be explored and under the context of adversarial training
for a target generative task \cite{Adate2019}.

In this paper, we design a novel RNN-based model with LSTM to simultaneously
learn a discriminative and generative task by utilizing the state
information in a shared LSTM layer. Using fMRI ROI time-series as
inputs, we interpret the LSTM block as modeling the coordination of
functional activity in the brain and the nodes of the LSTM as representing
functional communities, i.e., groupings of the input brain ROIs that
work together to both generate the fMRI time-series and perform the
discriminative task. We apply the proposed network for classification
of ASD vs. healthy controls, validating on multiple datasets from
the Autism Brain Imaging Data Exchange (ABIDE) I dataset. Compared
to several recent methods, we achieve some of the highest accuracy
reported on single-site ABIDE data. Finally, we evaluate the generative
results by analyzing the robustness of the extracted functional communities
and validate influential communities for classification in the context
of ASD.

\section{Methods}

\subsection{Network Architecture}

\subsubsection*{LSTM Block for Communities}

The LSTM module was designed to learn long-term dependencies in sequential
data \cite{Hochreiter1997}. An LSTM cell is composed of 4 neural
network layers with $K$ nodes that modulate two state vectors, the
hidden state $h_{t}\in\mathbb{\mathbb{R}}^{K}$ and the cell state
$c_{t}\in\mathbb{\mathbb{R}}^{K}$. The state vectors are updated
using input from the current time point $x_{t}\in\mathbb{R}^{R}$
and state information from the previous time point $h_{t-1}$ and
$c_{t-1}$: 
\begin{eqnarray}
g_{t} & = & \sigma\left(W_{g}x_{t}+U_{g}h_{t-1}+b_{g}\right),\textrm{with }g\in\left\{ i,f,o\right\} \\
\tilde{c_{t}} & = & \tanh\left(W_{c}x_{t}+U_{c}h_{t-1}+b_{c}\right)\\
c_{t} & = & i_{t}\ast\tilde{c_{t}}+f_{t}\ast c_{t-1},\quad h_{t}=o_{t}\ast\tanh\left(c_{t}\right)
\end{eqnarray}
where for layer $l\in\left\{ i,f,o,c\right\} $, $W_{l}$ are the
weights for the input, $U_{l}$ are the weights for the hidden state,
and $b_{l}$ are the bias parameters. 

The proposed network first takes the fMRI ROI time-series as inputs
to an LSTM layer (Fig. \ref{fig:Network}, blue path). The purpose
of this layer is to discover meaningful groupings of the ROIs, i.e.
functional communities, that are important for both generating and
classifying the input data. The LSTM block acts as a model for the
interaction between $R$ individual ROIs and $K$ functional communities
formed by the brain network to generate community activity. The activity
generated by each functional community $k$ is then represented by
the hidden state $h_{t}\left(k\right)$ and cell state $c_{t}\left(k\right)$,
which will serve as inputs to the rest of the network. 

Standard community detection methods for fMRI perform clustering based
on functional connectivity, where highly positively correlated ROIs
are grouped into a community. In our approach, we propose defining
a functional community by the interactions modeled in the LSTM and
the generated ROI data (see Sec. \ref{sub:FC}). To ensure that ROIs
within a community have positive ties as in standard approaches, we
constrain the input weights $W_{l}$ to be non-negative.

\subsubsection*{Discriminative Path}

The discriminative portion of the network aims to classify subjects
with ASD vs. typical controls (Fig. \ref{fig:Network}, orange path).
The architecture is similar to the classification network proposed
in \cite{Dvornek2017}. The difference is our approach first processes
the input time-series through an LSTM layer that learns to represent
functional communities of the ROI data. The \emph{hidden state} of
the LSTM cell at each time point is then fed to another LSTM layer,
followed by a shared dense layer with a single node, a mean pooling
layer, and a sigmoid activation to give the probability of ASD.

\subsubsection*{Generative Path}

\begin{figure}
\centering{}\includegraphics[bb=0bp 100bp 792bp 467bp,clip,width=0.78\textwidth]{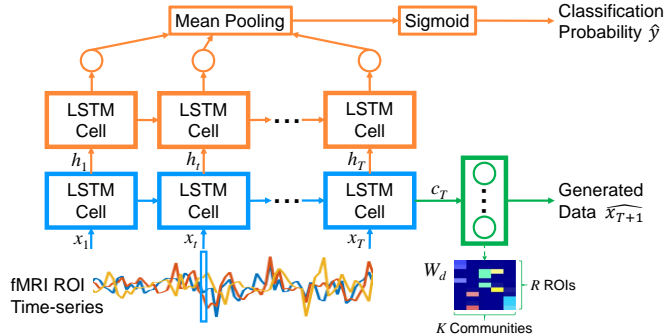}\caption{\label{fig:Network}Architecture of our jointly discriminative and
generative RNN: LSTM for functional communities (blue), discriminative
path (orange), and generative path (green).}
\end{figure}

The generative portion of the network looks to generate the data at
the next time point $x_{T+1}$ of an input time-series with length
$T$ (Fig. \ref{fig:Network}, green path). The input is first processed
by the same LSTM layer for functional communites as in the discriminative
network. The final \emph{cell state} $c_{T}$ of the LSTM cell is
then passed to a dense layer with $R$ nodes to produce the predicted
ROI values for the next time point $\widehat{x_{T+1}}=W_{d}c_{T}+b_{d}$.
To enforce that communities exert a positive influence on their members,
we constrain the weights for this dense layer to be non-negative.

\subsubsection*{Model Training}

The discriminative and generative paths are tied together during training
with the loss function $L=L_{G}\left(x_{T+1},\widehat{x_{T+1}}\right)+\lambda L_{D}\left(y,\hat{y}\right)$,
where $L_{G}$ is the loss for the generative model, $L_{D}$ is the
loss for the discriminative model, $y\in\left\{ 0,1\right\} $ is
the true label (1 denoting ASD), $\hat{y}$ is the predicted probability
of ASD, and $\lambda$ is a hyperparameter to balance the two losses.
For regularization, we include dropout layers before the shared dense
layer and mean pooling layer in the discriminative network and before
the dense layer in the generative network.

\subsection{Extraction of Functional Communities \label{sub:FC}}

As described above, we propose interpreting each node of the first
LSTM block as representing a functional community, where community
activity is summarized by state vectors $h_{t}$ and $c_{t}$. Since
it is difficult to analyze the interactions between ROIs and communities
via all the layers of the LSTM block, we propose defining the communities
based on their influence on each individual ROI. Recall that the generative
path uses the cell state $c_{T}$ as input to a dense layer to generate
the next ROI values $\widehat{x_{T+1}}=W_{d}c_{T}+b_{d}$. From a
graph structure perspective, a community is defined by densely connected
nodes, i.e. each member of a community is strongly influenced by that
community, but also the community is strongly influenced by its members.
Thus, we will use the weights $W_{d}\in\mathbb{R}^{R\times K}$ to
denote the membership between individual ROIs and their functional
communities. Row $r$ of $W_{d}$ represents the influence of each
community on ROI $r$, while column $k$ of $W_{d}$ represents the
influence of each ROI on community $k$. To provide hard membership
assignments, we perform k-means clustering with 2 clusters on the
membership weights in column $k$ of $W_{d}$ and assign the extracted
ROIs in the cluster with larger weights to community $k$ (Fig. \ref{fig:Network},
lower right).

\section{Experiments}

\subsection{Data}

We used resting-state fMRI data from the four ABIDE I \cite{DiMartino2014}
sites with the largest sample sizes: New York University (NY), University
of Michigan (UM), University of Utah School of Medicine (US), and
University of California, Los Angeles (UC). We selected preprocessed
data from the Preprocessed Connectomes Project \cite{Craddock2013}
using the Connectome Computation System pipeline, global signal regression
and band-pass filtering, and the Automated Anatomical Labeling (AAL)
parcellation with 116 ROIs. The extracted mean time-series of each
ROI was standardized (subtracted mean, divided by standard deviation)
for each subject.

Since the number of subjects per site is small for neural network
training, we augmented the datasets by extracting all possible consecutive
subsequences with length $T=30$ (i.e., 1 min. scantime) from each
subject, producing inputs of size $30\times116$. Thus, we augmented
the data by a factor of \textasciitilde{}150-250 for a total of \textasciitilde{}14000-38000
samples per site. At test time, the predicted probability of ASD for
a given subject was set to the proportion of subsequences labeled
as ASD.

\subsection{Experimental Methods}

Models for classification of ASD vs. control were trained for each
individual ABIDE site. We implemented the following LSTM-based networks
which all take the ROI time-series data as input: the proposed joint
discriminative/generative LSTM network (LSTM-DG); the same network
but using the hidden state for both data generation and class discrimination
(LSTM-H); the same network but with no generative constraint, i.e.
only the discriminative loss (LSTM-D); and a single layer discriminative
LSTM network as proposed in \cite{Dvornek2017} (LSTM-S). Models were
implemented in Keras, with 50 nodes for the first LSTM (for functional
communities) and 20 nodes for the second LSTM. Optimization was performed
using the Adam optimizer, with binary cross-entropy for $L_{D}$,
mean squared error for $L_{G}$, a batch size of 32, and early stopping
based on validation loss and a patience of 10 epochs. For joint discriminative/generative
networks, we set $\lambda=0.1$ so that $L_{G}$ and $L_{D}$ are
on similar scales. We also implemented a traditional learning pipeline
for resting-state fMRI (FC-SVM) \cite{Abraham2017}: the functional
connectivity based on Pearson correlation was input to a linear support
vector machine with L2 regularization, using nested cross-validation
to choose the penalty hyperparameter. All implemented models were
trained and tested on the augmented datasets. In addition, we compared
published results for the same ABIDE datasets and AAL atlas, including
another time-series modeling approach using hidden markov models (HMM)
\cite{Jun2019} and another neural network approach based on stacked
autoencoders and deep transfer learning (DTL) \cite{Li2018}.

To assess our implemented models, we used 10-fold cross-validation
(CV), keeping all data from the same subject within the same partition
(training, validation, or test). We measured model classification
performance by computing the accuracy (ACC), true positive rate (TPR),
true negative rate (TNR), and area under the receiver operating characteristic
curve (AUC). Paired one-tailed t-tests were used to compare model
performance over all folds and datasets.

For the generative results, with no ground truth for functional communities,
we instead evaluated the robustness of extracted communities and compared
a tensor decomposition approach for finding overlapping communities.
For each sample, we calculated the correlation matrix of the $R$
ROI time-series, then generated a tensor $\mathbf{T}$ with dimension
$R\times R\times S$, where $S$ is the number of samples. We then
used non-negative PARAFAC \cite{Carroll1970} to decompose $\mathbf{T}\approx\sum_{k=1}^{K}a_{k}\circ b_{k}\circ c_{k}$,
where $K$ is the number of communities, $a_{k}=b_{k}\in\mathbb{R}^{R}$
contains the membership weight of each ROI to community $k$, $c_{k}\in\mathbb{R}^{S}$
contains the membership weight of each sample to community $k$, and
$\circ$ is the vector outer product. Similar to our approach, we
set $K=50$ communities and use k-means clustering to assign hard
ROI memberships to each community. Then for each approach, we computed
the correlation of the membership weights and the Dice similarity
coefficient (DSC) of hard membership assigments between community
$k$ in fold 1 and all communities in fold $f\neq1$. The robustness
of community $k$ in fold 1 compared to fold $f$ was measured as
the maximum correlation/DSC computed in fold $f$. We then assessed
overall community robustness between fold 1 and $f$ using the average
correlation/DSC over all communities.

We also performed validation of the functional communities in the
context of the ASD classification task using Neurosynth \cite{Yarkoni2011},
which correlates over 14000 fMRI studies with 1300 descriptors. The
influence of a community for classification was denoted by the sum
of absolute weights across all nodes in the second LSTM block for
the discriminative task. A binary mask of the extracted ROIs for an
important discriminative community was then input to Neurosynth to
assess neurocognitive processes associated with ASD classification.

\subsection{Classification Results}

Classification results for each ABIDE site are in Tables \ref{tab:NY-UM-Results}
and \ref{tab:US-UC-Results}. Our LSTM-DG model produced the highest
accuracy for 3 of the 4 sites and second highest for US, in which
the LSTM-H variation of our model (generative path from hidden state)
performed best. Furthermore, LSTM-DG produced the highest or nearly
highest AUC for each site. Overall, our LSTM-DG consistently outperformed
all non-generative implemented models (ACC $p<0.05$) and showed potential
for improved classification compared to LSTM-H (ACC $p=0.08$). Moreover,
LSTM-DG was the only method to significantly outperform LSTM-S (ACC
$p=0.04$, TNR $p=0.04$), the original LSTM model for fMRI classification.
The results demonstrate the effectiveness of our proposed LSTM-DG
method to improve classification by jointly learning the generative
fMRI time-series model.
\begin{table}[b]
\centering{}\caption{\label{tab:NY-UM-Results}NY and UM Classification Results}
\resizebox{0.999\columnwidth}{!}{%
\begin{tabular}{|c||c|c|c|c||c|c|c|c|}
\hline 
 & \multicolumn{4}{c||}{NY (184 subjects, 42.3\% ASD)} & \multicolumn{4}{c|}{UM (143 subjects, 46.2\% ASD)}\tabularnewline
\cline{2-9} 
Model & ~Mean (Std)~ & ~Mean (Std) ~ & ~Mean (Std)~ & \multirow{2}{*}{~AUC~} & ~Mean (Std)~ & ~Mean (Std) ~ & ~Mean (Std)~ & \multirow{2}{*}{~AUC~}\tabularnewline
 & ACC (\%) & TPR (\%) &  TNR (\%) &  & ACC (\%) & TPR (\%) &  TNR (\%) & \tabularnewline
\hline 
LSTM-S \cite{Dvornek2017} & 69.5 (11.0) & 52.4 (26.5) & 83.1 (12.0) & 0.720 & 69.8 (11.4) & 56.7 (24.2) & 74.0 (25.3) & 0.740\tabularnewline
\hline 
FC-SVM \cite{Abraham2017} & 70.7 ~(8.2) & 54.8 (21.5) & 83.2 (11.8) & 0.783 & 69.2 (12.0) & 46.7 (18.9) & 89.8 (12.8) & 0.713\tabularnewline
\hline 
HMM \cite{Jun2019} & 70.6 ~(6.6) & 61.6~~~~~~~~\, & 66.7~~~~~~~~\, & 0.712 & 73.4 (10.5) & 68.5~~~~~~~~\, & 76.9~~~~~~~~\, & 0.738\tabularnewline
\hline 
DTL \cite{Li2018} & - & - & - & - & 67.2~~~~~~~~\, & 68.9~~~~~~~~\, & 67.6~~~~~~~~\, & 0.67~\,\tabularnewline
\hline 
\hline 
LSTM-D & 70.7 (11.0) & 48.9 (27.1) & 86.7 (16.1) & 0.746 & 67.0 (12.0) & 52.9 (22.2) & 78.6 (25.6) & 0.738\tabularnewline
\hline 
LSTM-H & 68.0~ (7.7) & 52.0 (19.8) & 80.1 (10.1) & 0.779 & 69.2 (11.4) & 57.9 (14.5) & 78.7 (18.1) & 0.777\tabularnewline
\hline 
\textbf{LSTM-DG} & \textbf{72.2 (14.7)} & \textbf{57.4 (25.5)} & \textbf{84.1 (12.2)} & \textbf{0.772} & \textbf{74.8 (10.0)} & \textbf{60.8 (12.8)} & \textbf{85.6 (14.5)} & \textbf{0.774}\tabularnewline
\hline 
\end{tabular}}
\end{table}
\begin{table}
\centering{}\caption{\label{tab:US-UC-Results}US and UC Classification Results}
\resizebox{0.999\columnwidth}{!}{%
\begin{tabular}{|c||c|c|c|c||c|c|c|c|}
\hline 
 & \multicolumn{4}{c||}{US (101 subjects, 57.4\% ASD)} & \multicolumn{4}{c|}{UC (99 subjects, 54.6\% ASD)}\tabularnewline
\cline{2-9} 
Model & ~Mean (Std)~ & ~Mean (Std) ~ & ~Mean (Std)~ & \multirow{2}{*}{~AUC~} & ~Mean (Std)~ & ~Mean (Std) ~ & ~Mean (Std)~ & \multirow{2}{*}{~AUC~}\tabularnewline
 & ACC (\%) & TPR (\%) &  TNR (\%) &  & ACC (\%) & TPR (\%) &  TNR (\%) & \tabularnewline
\hline 
\hline 
LSTM-S \cite{Dvornek2017} & 67.5 (15.4) & 79.8 (25.3) & 56.2 (41.8) & 0.659 & 62.7 (14.8) & 74.4 (31.5) & 51.5 (32.5) & 0.691\tabularnewline
\hline 
FC-SVM \cite{Abraham2017} & 67.3 (13.5)  & 86.2 (13.6) & 43.5 (27.6) & 0.721 & 61.7 (18.0) & 73.3 (20.6) & 47.7 (31.7) & 0.624\tabularnewline
\hline 
DTL \cite{Li2018} & 70.4~~~~~~~~\, & 72.5~~~~~~~~\, & 67.0~~~~~~~~\, & 0.73~\, & 62.3~~~~~~~~\, & 55.9~~~~~~~~\, & 68.0~~~~~~~~\, & 0.60~\,\tabularnewline
\hline 
\hline 
LSTM-D & 64.7 (17.8) & 75.3 (32.2) & 61.8 (39.6) & 0.682 & 63.6~ (8.8) & 71.8 (27.3) & 51.3 (30.5) & 0.662\tabularnewline
\hline 
LSTM-H & 76.4 (13.9) & 85.6 (18.0) & 65.8 (22.2) & 0.757 & 61.6 (11.4) & 66.6 (14.5) & 54.7 (18.1) & 0.705\tabularnewline
\hline 
\textbf{LSTM-DG} & \textbf{73.2 (14.7)} & \textbf{82.8 (25.5)} & \textbf{61.8 (12.2)} & \textbf{0.746} & \textbf{67.4 (10.0)} & \textbf{67.5 (12.8)} & \textbf{62.2 (14.5)} & \textbf{0.715}\tabularnewline
\hline 
\end{tabular}}
\end{table}

\subsection{Learned Functional Communities}

\begin{figure}[b]
\includegraphics[bb=250bp 250bp 550bp 390bp,clip,width=0.33\textwidth]{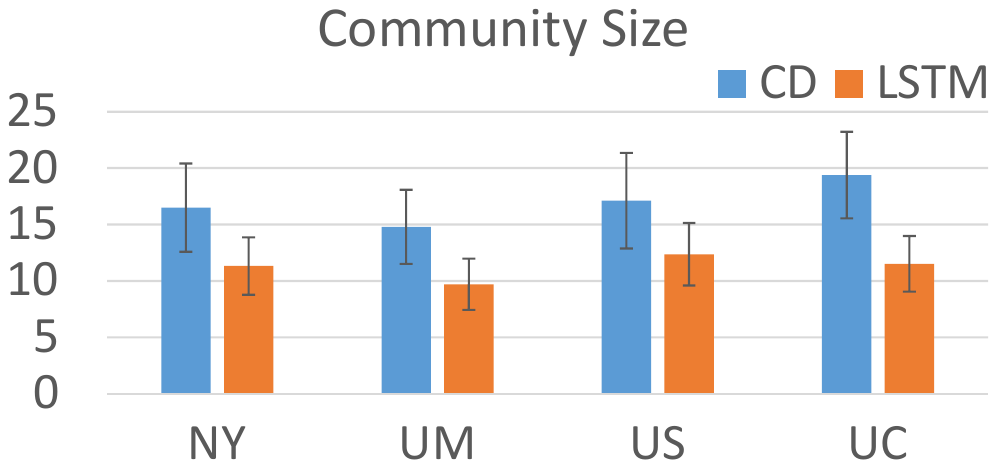}\hfill{}\includegraphics[bb=250bp 250bp 550bp 390bp,clip,width=0.33\textwidth]{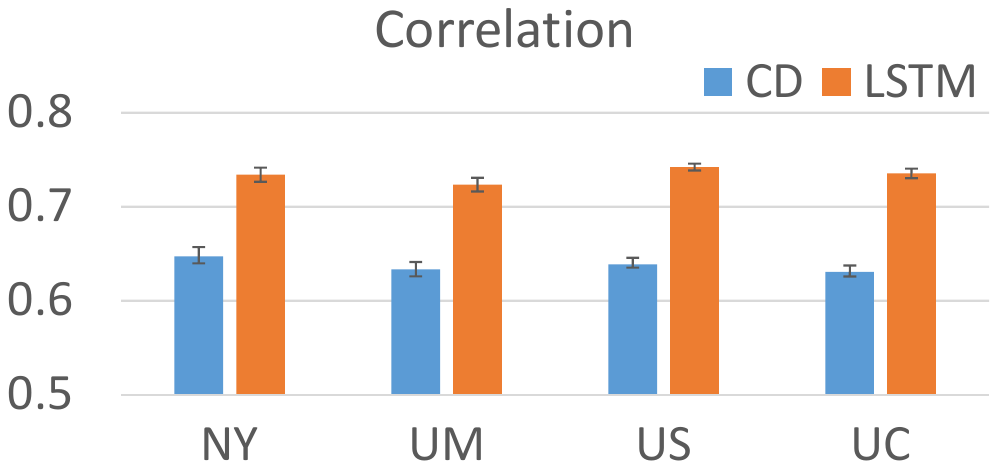}\hfill{}\includegraphics[bb=250bp 250bp 550bp 390bp,clip,width=0.33\textwidth]{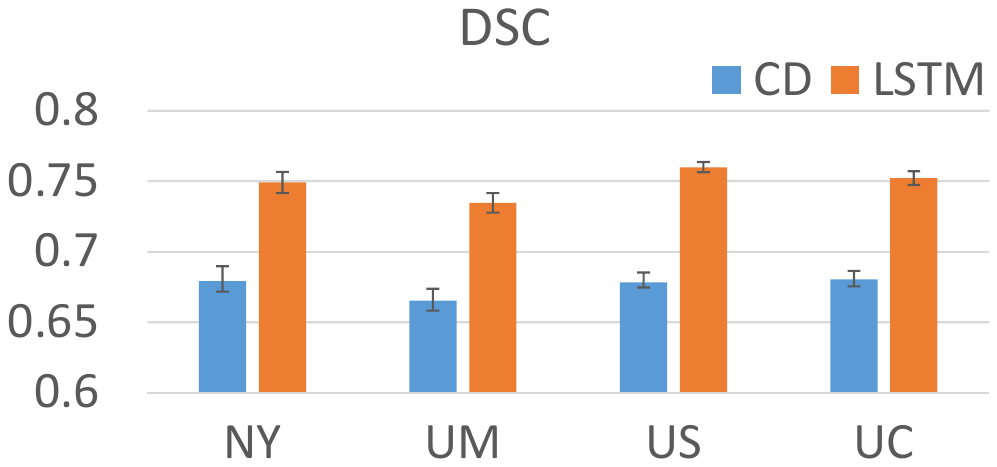}

\caption{\label{fig:robustness}Size (left) and robustness of extracted functional
communities across CV folds measured by correlation of membership
weights (middle) and DSC of hard assignments (right). CD = tensor-based
community detection, LSTM = proposed network.}
\end{figure}

Results for extracted communities by tensor-based community detection
(CD, blue) and the proposed LSTM approach (orange) are plotted in
Fig. \ref{fig:robustness}. Our LSTM method produced consistently
smaller communities with more uniform size compared to CD, with an
average of 11 ROIs compared to 16. Furthermore, our LSTM approach
consistently generated communities with higher correlation of membership
weights and higher DSC of hard community assignments across CV folds
for all sites, with a 15\% increase in average correlation and 11\%
increase in average DSC. Thus, our proposed network produced smaller
and more robust functional communities than CD, giving our model the
potential for more reliable interpretation of further analyses on
the functional communities. 

\begin{figure}[t]
\begin{centering}
\includegraphics[width=0.28\textwidth]{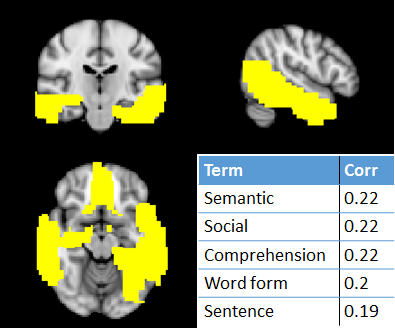}\hfill{}\includegraphics[width=0.28\textwidth]{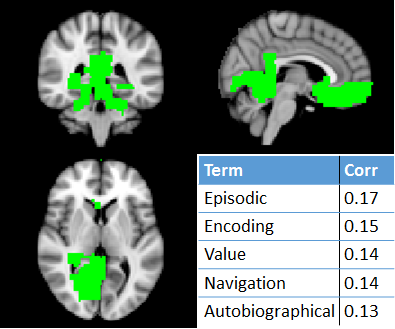}\hfill{}\includegraphics[width=0.28\textwidth]{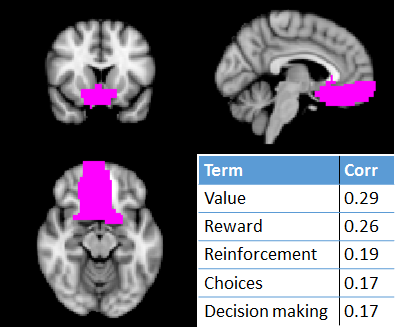}
\par\end{centering}

\caption{\label{fig:neurosynth}Top 3 influential communities for ASD classification
of the NY dataset and the top associated neurocognitive terms from
Neurosynth. }
\end{figure}

The top 3 influential communities for the ASD classification of the
largest dataset (NY) were extracted from the best CV fold and analyzed
in Neurosynth. ASD is characterized by impaired social skills and
communiciation; thus, we expect to find communities related to associated
neurological functions. The top extracted community (Fig. \ref{fig:neurosynth},
yellow) includes the temporal lobe and ventromedial prefrontal cortex,
which are associated with social and language processes. The second
community (Fig. \ref{fig:neurosynth}, green) includes the ventromedial
prefrontal cortex, hippocampus, and amygdala, which are associated
with memory. The third community (Fig. \ref{fig:neurosynth}, pink),
containing the ventromedial prefrontal cortex and ventral striatum,
is involved in reward processing and decision making. Dysfunction
of all these brain regions and processes in ASD have previously been
shown \cite{Kaiser2010}. 

\section{Conclusions}

We have presented a novel RNN-based network for jointly learning a
discriminative task and a generative model for fMRI time-series data.
We achieved higher ASD classification performance on several datasets,
demonstrating the advantage of joint learning. Finally, we showed
that functional communities defined by the LSTM nodes provide robust
representations of brain activity and facilitate interpretation of
the ASD classification model. Understanding functional network organization
will offer insights into brain disease as well as healthy cognition.

\bibliographystyle{splncs04}
\bibliography{miccai2019_ref}

\begin{thebibliography}{10}
\providecommand{\url}[1]{\texttt{#1}}
\providecommand{\urlprefix}{URL }
\providecommand{\doi}[1]{https://doi.org/#1}

\bibitem{Abraham2017}
Abraham, A., Milham, M.P., Martino, A.D., Craddock, R.C., Samaras, D., Thirion,
  B., Varoquaux, G.: Deriving reproducible biomarkers from multi-site
  resting-state data: An autism-based example. Neuroimage  \textbf{147},
  736--745 (2017)

\bibitem{Adate2019}
Adate, A., Tripathy, B.: S-lstm-gan: Shared recurrent neural networks with
  adversarial training. In: 2nd International Conference on Data Engineering
  and Communication Technology (2019)

\bibitem{Carroll1970}
Carroll, J., Chang, J.: Analysis of individual differences in
  multidimensionalscaling via an n-way generalization of eckart-young
  decomposition. Psychometrika  (1970)

\bibitem{Caruana1997}
Caruana, R.: Multitask learning. Machine Learning  (1997)

\bibitem{Craddock2013}
Craddock, C., Benhajali, Y., Chu, C., Chouinard, F., Evans, A., Jakab, A., ...,
  Bellec, P.: The neuro bureau preprocessing initiative: open sharing of
  preprocessed neuroimaging data and derivatives. In: Neuroinformatics (2013)

\bibitem{DiMartino2014}
Di~Martino, A., Yan, C.G., Li, Q., Denio, E., Castellanos, F.X., Alaerts, K.,
  ..., Milham, M.P.: The autism brain imaging data exchange: towards a
  large-scale evaluation of the intrinsic brain architecture in autism.
  Molecular Psychiatry  (2014)

\bibitem{Dvornek2017}
Dvornek, N.C., Ventola, P., Pelphrey, K.A., Duncan, J.S.: Identifying autism
  from resting-state fmri using long short-term memory networks. In: MLMI 2017.
  LNCS 10541 (2017)

\bibitem{Gueclue2017}
Güçlü, U., van Gerven, M.A.J.: Modeling the dynamics of human brain activity
  with recurrent neural networks. Front Comput Neurosci  (2017)

\bibitem{Graves2014}
Graves, A.: Generating sequences with recurrent neural networks.
  https://arxiv.org/abs/1308.0850 (2014)

\bibitem{Hochreiter1997}
Hochreiter, S., Schmidhuber, J.: Long short-term memory. Neural Computation
  (1997)

\bibitem{Jun2019}
Jun, E., Kang, E., Choi, J., Suk, H.I.: Modeling regional dynamics in
  low-frequency fluctuation and its application to autism spectrum disorder
  diagnosis. NeuroImage  (2019)

\bibitem{Kaiser2010}
Kaiser, M., Hudac, C., Shultz, S., Lee, S., Cheung, C., Berken, A., ...,
  Pelphrey, K.: Neural signatures of autism. Proc Natl Acad Sci U S A  (2010)

\bibitem{Li2018}
Li, H., Parikh, N.A., He, L.: A novel transfer learning approach to enhance
  deep neural network classification of brain functional connectomes. Front.
  Neurosci.  (2018)

\bibitem{Li2018a}
Li, H., Fan, Y.: Brain decoding from functional mri using long short-term
  memory recurrent neural networks. MICCAI 2018  (2018)

\bibitem{Yarkoni2011}
Yarkoni, T., Poldrack, R.A., Nichols, T.E., {Van Essen}, D.C., Wager, T.D.:
  Large-scale automated synthesis of human functional neuroimaging data. Nature
  Methods  (2011), www.neurosynth.org

\end{thebibliography}

\end{document}